\documentclass[aps,prd,nofootinbib,fleqn,showpacs,preprint,groupedaddress]{revtex4}
\usepackage{graphicx}
\usepackage{amssymb}
\usepackage{bm}

\begin{document}

\title
{Light-front wavefunction dependence of the quark recombination}



\author{Byungsik Hong}
\email{bhong@korea.ac.kr}
\affiliation{Department of Physics, Korea University, Seoul 136-701, Korea}
\author{Chueng-Ryong Ji}
\email{ji@ncsu.edu}
\affiliation{Department of Physics, North Carolina State University, Raleigh, North Carolina 27695}
\author{Dong-Pil Min}
\email{dpmin@snu.ac.kr}
\affiliation{School of Physics, Seoul National University, Seoul 151-747, Korea}


\date{\today}

\begin{abstract}
We present an extension of the recombination formalism to 
analyze the effects from the variation of the hadron wavefunctions.
The hadron spectra are sensitive to the shape of the wavefunctions.
However, when we fit the wavefunction parameters to the physical 
observables, such as the average charge radius, 
the final spectra are very similar each other. 
We discuss our numerical results in comparison with 
the published PHENIX and STAR data at RHIC. 
In the hadron spectra, the recombination of thermal partons dominates 
at intermediate transverse momentum ($P_{T}$ = 2 $\sim$ 5 GeV), 
and the fragmentation dominates at high $P_{T}$ ($>$ 5 GeV).
The yield ratios and the nuclear modification factors for various hadron 
species are also estimated and compared to the experimental data. 
We present a new prediction on $\bar{p}/p$ and $K^{-}/K^{+}$ ratios, 
including the jet quenching effects to the fragmentation mechanism. 
\end{abstract}



\maketitle

\section{Introduction}\label{introduction}

Although no one doubt the existence of quarks and gluons, they have not 
yet been detected individually at the zero temperature. The current 
vigorous efforts at the Relativistic Heavy Ion Collider (RHIC) 
and the future plans of the Large Hadron Collider (LHC) may reveal 
the temperature dependence of the confinement mechanism and 
the chrial symmetry restoration \cite{lee1}. 
The high-energy nuclear collisions compress and heat the heavy nuclei 
so much that their individual protons and neutrons may overlap and, 
in addition, a lot of pions may arise to ultimately 
create the quark-gluon plasma (QGP). 
The QGP may have existed ten millionths of second after the big 
bang and created the primordial matter of the universe. 
The RHIC and the future LHC may yield the QGP in the laboratory. 
It has been reported that the four experiments at RHIC already 
obtained distinguished results from the lower energy
heavy ion collisions at CERN SPS \cite{b-wp,p-wp,ph-wp,s-wp}. 
The future LHC experiments (ALICE as well as CMS and ATLAS)
would require theoretical predictions 
at the 30-fold energy increase from the RHIC. 

Among many others, the effects from the jet quenching and 
the bulk hadronization may be regarded as the important new results. 
Especially, the elliptic flow analysis revealed that 
the differential second-harmonic Fourier moment ($v_{2}$) 
of the azimuthal distribution with respect to 
the reaction plane had a remarkable saturation property 
in the intermediate transverse momentum ($P_{T}$) range 
between 2 and 6 GeV for all hadrons including multistrange baryons.
This saturation effect and eventual decrease of $v_{2}$ at high $P_{T}$ 
have been qualitatively interpreted to be the results of 
partonic energy loss in an opaque parton system created by 
nuclear collisions \cite{wang-1,wang-2}. 
Furthermore, the estimated $v_{2}$ parameter as a function of $P_{T}$ 
are scaled by the number of constituent quarks of particles. 
Together with an enhanced proton production in 
the intermediate $P_{T}$ region, this agreement 
has been taken seriously as one piece of evidence 
for the quark recombination process and the presence of 
partonic collectivity at the early stage 
of a collision \cite{fries,mola-1,grec-1,qc-1,qc-2,hong-1}.

In this work, we utilize the previous recombination formalism, 
and extend it to analyze the light-front (LF) wavefunction dependence 
in the theoretical predictions from this formalism \cite{fries}. 
Typical forms of the LF wavefunctions such as the Gaussian form 
and the power-law form are applied to this extended formulation. 
The numerical results are contrasted to each other, 
and compared with the available single invariant spectra 
by PHENIX and STAR for various mesons and baryons. 
We also discuss the production ratios of various hadrons, 
including ${\bar p}/p$ and $K^-/K^+$, in the fragmentation region. 
While we include the jet quenching effects as others do, 
we get a rather distinguished results compared to the previous ones. 
For the high $P_T$ regions, we get a dramatic suppression of 
the antiparticles $\bar p$ and $K^-$ compare to 
the corresponding particles $p$ and $K^+$, respectively.

The paper is organized as follows. 
In Sec.~\ref{formulation}, we present the recombination formalism 
which is extended from the previous one to explicitly 
include the intrinsic transverse momenta of the constituents 
inside the hadron. 
Rather than an extensive review of the previous formalism, 
we focus on what has been extended from the previous model. 
In Sec.~\ref{result}, we present the numerical results of 
the $P_T$ spectra for various mesons and baryons to contrast 
the results between the Gaussian form and the power-law form. 
The results are compared with the available experimental data 
from PHENIX and STAR collaborations.
We also discuss the production ratios of various hadrons 
and the nuclear modification factor $R_{CP}$ in this section. 
Conclusions and discussions follow in Sec.~\ref{conc}.

\section{Formulation}\label{formulation}

\subsection{Recombination and Light-Front Wavefunction Dependence}
\label{recombi}

The current data from the RHIC experiments seem to indicate 
two distinguished mechanisms of hadronization: 
(1) quark recombination for a rather low and intermediate $P_T$ region and 
(2) quark fragmentation for a high $P_T$ region. 
In this section, we present an extension of the recombination formalism 
to analyze the effects from the variation of the hadron wavefunctions. 

Introducing the density matrix $\hat\rho$ for the system of partons, 
the number of quark-antiquark states that one may interprete 
as mesons is given by
\begin{equation}
N_M = \Sigma_{a b} \int \frac{d^3 P}{(2\pi)^3} <M;P|{\hat\rho}_{ab}|M;P>,
\label{number}
\end{equation}
where $|M;P>$ is a meson state with the momentum $P$ and the sum is over 
all combinations of quantum numbers such as flavor, helicity, and 
color of the valence quarks that contribute to the given meson $M$.
Inserting complete sets of coordinates and using 
the Wigner function formalism, one can derive the formula 
for the invariant spectrum of the meson $M$ as follows \cite{fries}: 
\begin{eqnarray}
E \frac{d^{3}N_M}{dP^{3}}&=&C_M \int_\Sigma \frac{d^3 R P\cdot u(R)}{(2\pi)^3}
\int \frac{d^3 q}{(2\pi)^3} 
\nonumber \\ &&\times 
w_a(R; \frac{P}{2}-q)
\Phi^W_M (q) w_b(R;\frac{P}{2}+q) \nonumber \\
&=& C_M \int_\Sigma \frac{d^3 R P\cdot u(R)}{(2\pi)^3} 
\int \frac{dx P^+ d^2 {\vec k}_\perp}{(2\pi)^3}
\nonumber \\ &&\times 
w_a(R; xP^+, {\vec k}_\perp) \Phi_M (x, {\vec k}_\perp) 
w_b(R;(1-x)P^+, -{\vec k}_\perp),
\label{spectrum-meson}
\end{eqnarray}
where $\Phi^W_M(q) = \int d^{3}r \Phi^W_M(r,q)$ 
in the Wigner function formalism and 
$\Phi_M(x,{\vec k}_\perp) = |\psi_M (x, {\vec k}_\perp)|^2$
using the LF wavefunction of the meson $\psi_M (x, {\vec k}_\perp)$. 
Here, $x$ and ${\vec k}_{\perp}$ are the momentum fraction and 
the respective intrinsic transverse momentum of each quark. 
Similarly, the invariant spectrum of the baryon $B$ 
can be obtained as follows \cite{fries}:
\begin{eqnarray}
E \frac{d^{3}N_B}{dP^{3}}&=& C_B \int_\Sigma 
\frac{d^3 R P\cdot u(R)}{(2\pi)^3} 
\int \frac{dx_{1}P^{+} d^{2}{\vec k}_{1\perp}}{(2\pi)^{3}} 
\int \frac{dx_{2}P^{+} d^{2}{\vec k}_{2\perp}}{(2\pi)^{3}} 
\int \frac{dx_{3}P^{+} d^{2}{\vec k}_{3\perp}}{(2\pi)^{3}} 
\nonumber \\ &&
\times 
w_a(R; x_{1} P^{+}, {\vec k}_{1\perp}) 
w_b(R; x_{2} P^{+}, {\vec k}_{2\perp})
w_c(R; x_{3} P^{+}, {\vec k}_{3\perp})
\nonumber \\ &&
\times \Phi_B (x_{1}, x_{2}, x_{3}, {
\vec k}_{1\perp}, {\vec k}_{2\perp},  {\vec k}_{3\perp}), 
\label{spectrum-baryon}
\end{eqnarray}
where $x_{1} + x_{2} + x_{3} = 1$ and 
${\vec k}_{1\perp}+{\vec k}_{2\perp}+{\vec k}_{3\perp} = 0$.

The previous work by Fries and collaborators used a factorized ansatz
for the LF wavefunction, for example \cite{fries}, 
\begin{equation}
\psi_M (x,{\vec k}_\perp) = \phi_M(x) \Omega({\vec k}_\perp),
\label{ansatz}
\end{equation}
for mesons with a longitudinal distribution amplitude $\phi_M (x)$ 
and a transverse part $\Omega({\vec k}_\perp)$. 
However, we note that such factorization ansatz cannot be justified 
in free space 
since the LF wavefunction depends on 
the LF invariant mass of the particle, e.g., for the meson 
$(m_a^2 + {\vec k}_\perp^2)/x + (m_b^2 + {\vec k}_\perp^2)/(1-x)$
(here the meson is composed of quark $a$ and $b$), which 
cannot be factorized as Eq.(\ref{ansatz}). 
In general, the assumption of wavefunction factorization 
such as Eq.(\ref{ansatz}) is not acceptable 
in free space 
because LF wavefunctions should be solutions of LF bound-state 
equations and the LF energy-momentum dispersion 
relation is rational, i.e.
$k^- = ({k_\perp}^2 + m^2)/{k^+}$ for the particle with mass $m$.
Both the LF kinetic energy (i.e. the LF invariant mass of the bound-state) 
and the LF kernel (or the inverse of the LF energy difference 
between the initial and intermediate states) involved in 
the LF bound-state equations are not factorizable due to 
the rational energy-momentum dispersion relations. 
Thus, the solutions of the LF bound-state equations cannot be factorizable
and we do not integrate over ${\vec k}_\perp$ in 
Eqs.(\ref{spectrum-meson}) and (\ref{spectrum-baryon}) 
but leave ${\vec k}_\perp$ explicitly in the formulation. 
On the other hand, since it is not yet known if the LF 
bound-state solution in free space is also applicable to the dynamical recombination 
process in quark matter without any modification, we note that 
the factorization ansatz used in Ref. \cite{fries} may be equally 
valuable as one of the model calculations in this work.

The usual parton spectrum at a given temperature is given by \cite{fries}
\begin{equation}
w_a (R; p) = \gamma_a e^{-p \cdot v(R)/T} \cdot 
e^{-\eta^2/{2 \Delta^2}} f(\rho,\phi),
\label{parton-spectrum}
\end{equation}
where $\rho$ and $\phi$ are the radial and 
the azimuthal angle coordinates, respectively. 
In addition, $v(R)$ and $\eta$ represent the velocity four vector 
and the rapidity of the quark $a$, respectively. 
Here, $\gamma_a = \exp(\mu_{a}/T)$ is the fugacity factor 
for each quark species $a$ for which we adopt the results from 
the statistical analysis for the hadron production at RHIC \cite{pbm-1}: 
the chemical potential $\mu_{a}$'s are 9, 6.7, and $-$3.9 MeV
for $a$ = $u$(or $d$), $s$, and $c$, respectively. 
Note that Ref. \cite{pbm-1} gives resulting chemical potentials 
for isospin, strangeness and charmness as well as 
baryon chemical potential estimated by the statistical model at RHIC. 
Since statistical analysis of hadron production provides 
only chemical potentials of hadrons (not quarks), 
we obtained the quark fugacities 
by using the following formula for the fugacity 
of hadron $i$ \cite{torri}:\footnote{The quark fugacity $\gamma_{a}$ 
in this paper means $\lambda_{a}$ in Eq.(3) of Ref. \cite{torri} 
with the saturation factor one.}

\begin{equation}
\Upsilon_{i} = \gamma_{I_{3}^{i}} \prod_{a} \gamma_{a}^{N_{a}^{i}}
\label{fugacity}
\end{equation} 
where $N_{a}^{i}$ is the number of quark species $a$ in hadron $i$. 
In Eq.(\ref{fugacity}), the fugacity $\gamma_{I_{3}^{i}}$ is close to 1, 
as the isospin chemical potential $\mu_{I_{3}^{i}}$ and 
the assumed freeze-out temperature $T$ 
are -0.96 MeV and 175 MeV, respectively \cite{pbm-1}. 
We assume that the temperature $T$ for hadronization occurs at 175 MeV. 
The lattice QCD predicts that the phase transition occurs 
at $T =$ 175 MeV at vanishing baryon chemical potential \cite{lqcd}. 
It should be reasonable that the temperature of the partonic phase 
is assumed to be the same as that of the phase transition. 
The space-time structure of the parton source in Eq.(\ref{parton-spectrum}) 
is given by a transverse distribution $f(\rho,\phi)$ and 
a wide Gaussian rapidity distribution with a width $\Delta$. 
Also, one may assume $f(\rho,\phi) \approx \Theta(\rho_0 -\rho)$ 
especially for the analysis of the central collisions.
With these assumptions, we find
\begin{eqnarray}
\frac{d^{3}N_M}{dP_{T}^{2}dy}|_{y=0}&=& C_M M_T \frac{V}{(2\pi)^3} 2 \gamma_a 
\gamma_b I_0 \left[\frac{P_T \sinh \eta_T}{T}\right] \nonumber \\
&&\times \int_0^1 dx \int_0^{\infty} d^2 {\vec k}_\perp 
|\psi_M (x, {\vec k}_\perp)|^2 k_M (x, 
{\vec k}_\perp, P_T),
\label{extend-spectrum}
\end{eqnarray}
where $V = \tau A_{T}$ ($\tau$ is the hadronization time and 
$A_{T}$ is the transverse size) is the volume of the parton system and 
\begin{equation}
k_M (x, {\vec k}_\perp, P_T) = K_1 \left[\frac{\cosh{\eta_T}}{T}
\{\sqrt{m_a^2 + (x P_T + {\vec k}_\perp)^2} + 
\sqrt{m_b^2 + \{(1-x) P_T - {\vec k}_\perp\}^2} \} \right].
\label{kM}
\end{equation}
We note that the particular combination of $P_T$ and ${\vec k}_\perp$
for each constituent quark in Eq.(\ref{kM}) is consistent with 
the boost invariance of $k_M$ in light-front dynamics. 
Extension to the baryon case is straightforward as
\begin{eqnarray}
\frac{d^{3}N_B}{dP_{T}^{2}dy}|_{y=0}&=& 
C_B M_T \frac{V}{(2\pi)^3} 2 \gamma_a \gamma_b \gamma_c 
I_0 \left[\frac{P_T \sinh \eta_T}{T}\right] 
\int_0^1 dx_{1} dx_{2} \int_0^{\infty} 
d^2 {\vec k}_{1\perp} d^2 {\vec k}_{2\perp}  
\nonumber \\
&&\times |\psi_B (x_{1},x_{2},
{\vec k}_{1\perp},{\vec k}_{2\perp})|^2 
k_B (x_{1}, x_{2}, {\vec k}_{1\perp}, {\vec k}_{2\perp}, P_T),
\label{Bextend-spectrum}
\end{eqnarray}
and
\begin{eqnarray}
k_B (x_{1}, x_{2}, {\vec k}_{1\perp}, {\vec k}_{2\perp}, P_T) &=& 
K_1 [ \frac{\cosh{\eta_T}}{T}
\{ \sqrt{m_a^2 + (x_{1} P_T + {\vec k}_{1\perp})^2}
 + \sqrt{m_b^2 + (x_{2} P_T + {\vec k}_{2\perp})^2}
\nonumber \\
&& +\sqrt{m_c^2 + \{(1-x_{1}-x_{2})P_T - ({\vec k}_{1\perp} + {\vec k}_{2\perp})\}^2}\} ].
\label{kB}
\end{eqnarray}
In the following analysis, we take $V$ as a free parameter to fit 
all invariant spectra of hadrons simultaneously 
for a given collision centrality. 

With this extension, we can now explicitly include the effect 
from the intrinsic transverse momentum of each quark, 
and vary the form of the LF wavefunction such as the Gaussian form 
and the power-law form \cite{ji-1}. 
In this analysis, we use the following typical LF wavefunctions 
for mesons and contrast the results between the two:
\begin{equation}
\psi_{Gaussian} (x, {\vec k}_\perp) = 
\exp \left[-(\frac{m_a^2 + {\vec k}_\perp^2}{x} + 
\frac{m_b^2 + {\vec k}_\perp^2}{1-x})/\beta^2 \right],
\label{Gaussian}
\end{equation}
and
\begin{equation}
\psi_{Power-law} (x, {\vec k}_\perp) = 
1/\left[\frac{m_a^2 + {\vec k}_\perp^2}{x} + 
\frac{m_b^2 + {\vec k}_\perp^2}{1-x} + \alpha^2 \right]^{n}, 
\label{power-law}
\end{equation}
where $\beta$, $\alpha$, and $n$ are the parameters 
that can be fixed from the physical observables such 
as the size and the mass spectrum of meson, etc.. 
The extension of Eqs.(\ref{Gaussian}) and (\ref{power-law}) to 
baryons is rather straightforward:
\begin{equation}
\psi_{Gaussian} (x_{1}, x_{2}, 
{\vec k}_{1\perp}, {\vec k}_{2\perp}) = 
\exp \left[-(\frac{m_a^2 + {\vec k}_{1\perp}^2}{x_{1}} + 
\frac{m_b^2 + {\vec k}_{2\perp}^2}{x_{2}} + 
\frac{m_c^2 + ({\vec k}_{1\perp} + {\vec k}_{2\perp})^2}
{1-x_{1}-x_{2}})/\beta^2 \right],
\label{B_Gaussian}
\end{equation}
and
\begin{equation}
\psi_{Power-law} (x_{1}, x_{2}, 
{\vec k}_{1\perp}, {\vec k}_{2\perp}) = 
1/\left[\frac{m_a^2 + {\vec k}_{1\perp}^2}{x_{1}} + 
\frac{m_b^2 + {\vec k}_{2\perp}^2}{x_{2}} + 
\frac{m_c^2 + ({\vec k}_{1\perp} + {\vec k}_{2\perp})^2}
{1-x_{1}-x_{2}} + \alpha^2 \right]^{n}.
\label{B_power-law}
\end{equation}

In this calculation, we used 260 MeV for the masses of 
$u$ and $d$ quarks and 460 MeV for the mass of $s$ quark. 
In a relativized quark model with chromodyamics, 
spectra of both mesons and baryons have been well analyzed. 
As shown in typical references 
(\cite{mass-m} for mesons and \cite{mass-b} for baryons), 
the potentials among constituents such as confinement, hyperfine,
spin-orbit, etc. work together to reproduce the hadron spectra 
comparable to the experimental values. 
For instance, proton mass was predicted as 960 MeV
and the mass difference between nucleon and delta was obtained 
around 300 MeV while the light quark mass was taken as 220 MeV. 
The same light quark mass was used to predict the meson spectra 
which were overall in good agreement with data. 
These support our light-front quark model calculations 
(see e.g. \cite{mass-3}). 
Although we took the light quark mass 260 MeV as used in Ref. \cite{fries}
for the present analysis, the essential predictions from
a relativized quark model (or light-front quark model) remain intact. 

\begin{figure}[t!]
\begin{center}
\includegraphics[width=16cm]{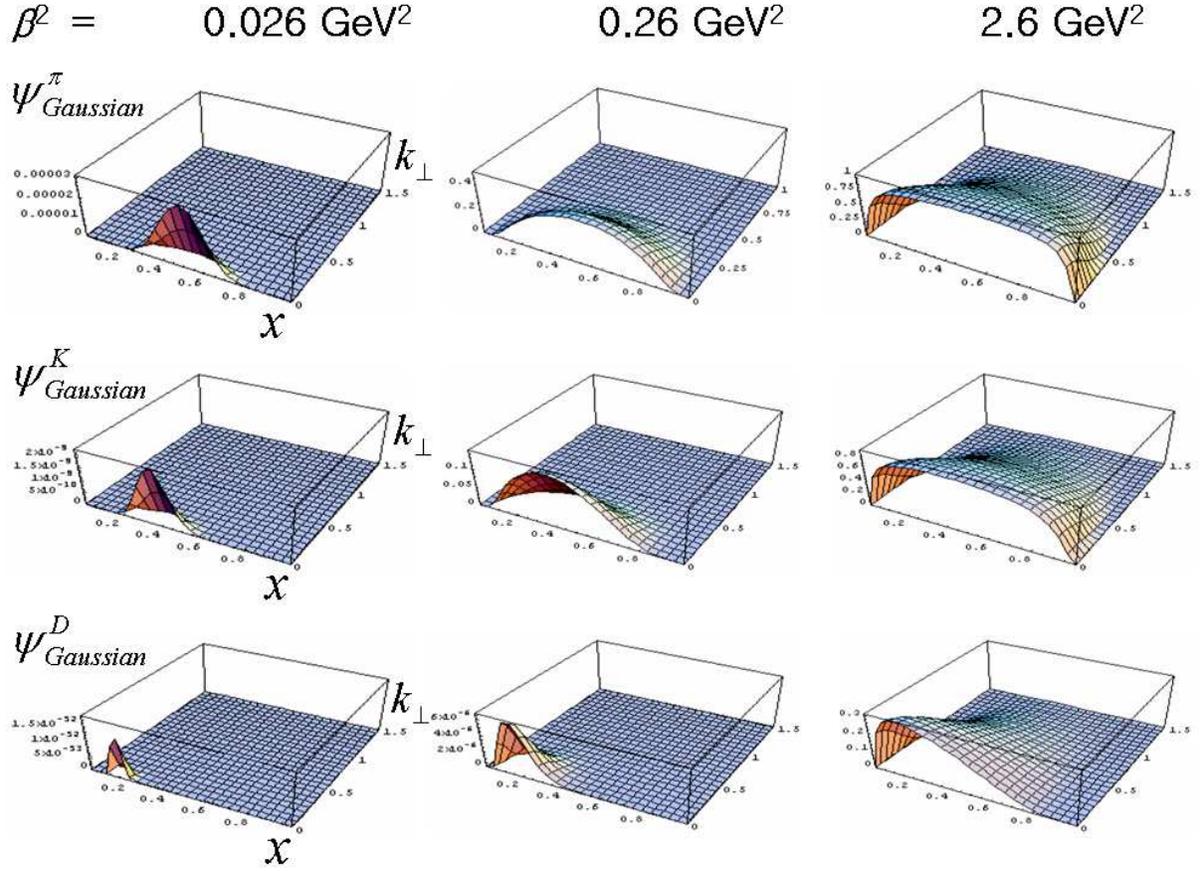}
\end{center}
\caption[]{Shapes of the Gaussian wavefunctions as functions of 
$x$ and $k_{\perp}$ for different $\beta^{2}$ 
(normalization is not performed in this figure).
The top, middle, and bottom rows represent the wavefunctions 
for $\pi$, $K$, and $D$, respectively.}
\label{Fig:wf_evolve}
\end{figure}

Just to illustrate how typical LF wavefunctions look like, 
we plot $\psi_{Gaussian} (x, {\vec k}_\perp)$ 
for different $\beta^{2}$ values in Fig.~\ref{Fig:wf_evolve}. 
As expected, the LF wavefunctions are symmetric around $x$ = 0.5, 
if the masses of constituent quark and antiquark are the same for pions. 
As $\beta^{2}$ increases, $\psi_{Gaussian} (x, {\vec k}_{\perp})$
becomes broader in $x$ as well as in ${\vec k}_\perp$. 
When the mass of constituent quark and antiquark 
are not the same like $K$ and $D$, 
$\psi_{Gaussian} (x, {\vec k}_{\perp})$'s are clearly skewed in $x$. 

\begin{figure}[t!]
\begin{center}
\includegraphics[width=13cm]{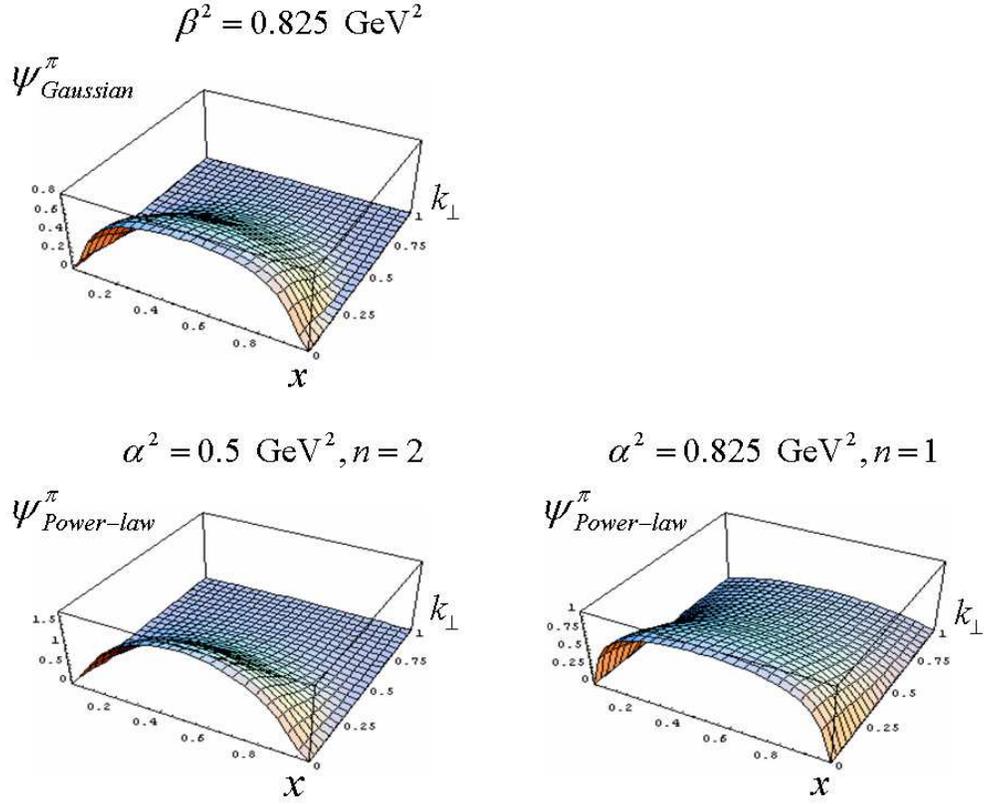}
\end{center}
\caption[]{Comparison of the pion wavefunctions between 
the Gaussian form and the power-law form.
The Gaussian wavefunction with $\beta^{2}$ = 0.825 GeV$^{2}$ (top) 
and the second order power-law wavefunction with 
$\alpha^{2}$ = 0.5 GeV$^{2}$ (bottom left) are adjusted 
to the average charge radius of pions.
But the first order power-law wavefunction with 
$\alpha^{2}$ = 0.825 GeV$^{2}$ (bottom right) is not adjusted.}
\label{Fig:wf_comp}
\end{figure}

In order to constrain $\beta$, $\alpha$, and $n$ 
in Eqs.(\ref{Gaussian}) - (\ref{B_power-law}), 
the average values of ${\vec k}_{\perp}$ are fixed by 
the measured average charge radius square $<r^{2}>$ of each hadron. 
If the experimental data for $<r^{2}>$ are not available, we adopt 
the calculated ones by a relativistic quark model \cite{ji-2,schl-1}.
As an example, $<r^{2}>$ = 0.44 fm$^{2}$ for pions, and 
the corresponding $\beta^{2}$ is 0.825 GeV$^2$ for $\psi_{Gaussian}$.
The average values of the charge radius square and 
the corresponding values of $\beta^{2}$ are summarized 
in Table~\ref{Tab:beta2}. 
In addition, the deduced $\alpha^{2}$ and $n$ of $\psi_{Power-law}$ 
are 0.5 (1.53) GeV$^2$ and 2 (6), respectively, for pions (protons). 
The left two panels of Fig.~\ref{Fig:wf_comp} show the comparison of 
$\psi_{Gaussian}$ and $\psi_{Power-law}$ of pions 
by using the adjusted wavefunction parameters. 
They demonstrate that the LF wavefunctions are very similar in shape 
when the parameters are determined by some physical observables 
such as the charge radii of hadrons. 
However, if we use some arbitrary values for those parameters, 
the shape of LF wavefunctions can be quite different, 
which is demonstrated in the lower right panel of Fig.~\ref{Fig:wf_comp}. 

\begin{figure}[t!]
\begin{center}
\includegraphics[width=12cm]{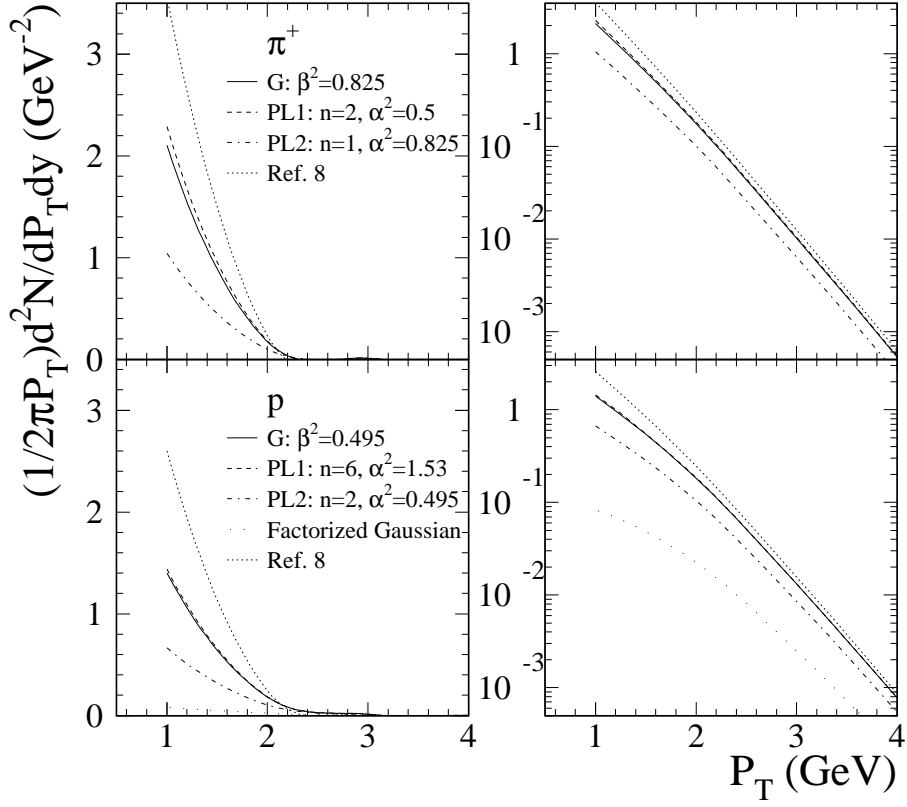}
\end{center}
\caption[]{Comparison of the invariant spectra of $\pi^{+}$ 
(top) and $p$ (bottom) by the recombination process for 
various assumptions on the wavefunction parameters. 
Left panels show the spectra in linear scale in order to 
emphasize the difference in relatively low $P_T$ region, 
while right panels show them in log scale 
for the comparison of the overall shapes. 
G represents the Gaussian wavefunction used in this paper. 
PL1 represents the power-law wavefunction whose 
parameters are adjusted by the known average charge radius, 
whereas PL2 represents the power-law wavefunction 
with arbitrary values for the parameters. 
The coarsely-dotted 
lines are calculated by a factorized Gaussian 
form of the LF wavefunctions with $\beta^2 =$ 0.495 GeV$^2$
(see text for details). 
For the comparison, the wavefunctions used in Ref.~\cite{fries} 
are also plotted by the dotted lines.
}
\label{Fig:wf_check}
\end{figure}

The importance of the proper choice of the LF wavefunction parameters
in the hadron spectra are explained in Fig.~\ref{Fig:wf_check} 
for $\pi$'s and protons. 
The invariant yields of the recombined hadrons are quite different
for different sets of parameters. However, the hadron yields 
from the recombination process are quite similar 
for $\psi_{Gaussian}$ and $\psi_{Power-law}$, 
once the wavefunction parameters are fixed by some physical observables
(See solid vs. dashed lines in Fig.~\ref{Fig:wf_check}.). 
In the following analysis, we use only the Gaussian wavefunction 
with proper $\beta^2$ for each hadron.
For the comparison, the wavefunctions used in Ref.~\cite{fries} 
are also included in Fig.~\ref{Fig:wf_check}. 
The wavefunctions used in the present analysis (solid lines) are 
lower than the factorized wavefunctions used in Ref.~\cite{fries} 
shown by dotted lines especially for relatively low $P_{T}$ region.
Even in the logarithmic scale, the differences are as visible as 
the ones with arbitrary wavefunction parameters (dash-dotted lines). 

Since the essence of this work is to study the effect of 
the proper treatment of the LF wavefunctions 
to the recombination yields, we also compare 
the invariant spectra estimated by using full LF wavefunctions 
with those estimated by factorized wavefunctions. 
For the approximation that uses factorized wavefunctions for mesons,
we tested $\psi_{Gaussian} (x, k_{\perp})$ of Eq.(\ref{Gaussian}) 
with $<{\vec k_{\perp}}^2> =$ 0.088 GeV$^2$, multiplied by 
$\Omega ({\vec k_{\perp}}) = \exp{(-{\vec k_{\perp}}^{2}/\beta^{2})}$
with $\beta^{2} =$ 0.176 GeV$^2$, which gave us 
the right $<{\vec k_{\perp}}^2>$ value. 
The resulting invariant spectrum for the recombined $\pi^{+}$'s 
is very similar to the PL1 option in Fig.~\ref{Fig:wf_check}. 
Since it duplicates the result of PL1 option, we do not display
it explicitly in Fig.~\ref{Fig:wf_check}. However, this result
indicates that the above factorized wavefunction may be 
as useful as the full LF wavefunctions developed in this paper 
for the phenomenological analysis of mesons, 
when the parameters are properly chosen. 

Similarly, for baryons, we tested 
$\psi_{Gaussian} (x_{1}, x_{2}, k_{1\perp}, k_{2\perp})$ 
of Eq.(\ref{B_Gaussian}) 
with $<{\vec k_{i\perp}}^2> =$ 0.0512 GeV$^2$, multiplied by 
$\Omega ({\vec k_{\perp}}) = 
\exp{[-({\vec k_{1\perp}}^{2}+{\vec k_{2\perp}}^{2})/2\beta^{2}]}$
with $\beta^{2} =$ 0.495 GeV$^2$. 
However, we could not find proper $\beta^2$ value 
for the factorized wavefunction to get the correct 
$<k_\perp^2>$ = 0.0512 GeV$^2$ since the results were 
not at all stable (too large standard deviations). 
Thus, we instead varied $\beta^2$ value in a wide range starting
from 0.03 GeV$^2$ all the way even above 1 GeV$^2$, 
and compared the resulting proton spectra. 
Results for $\beta^2$ below 0.03 GeV$^2$ couldn't be obtained
due to a numerical instability. The results for $\beta^2$ above
1 GeV$^2$ were about the same as the result of $\beta^2 =$ 1 GeV$^2$
and very stable as expected from the form of the above factorized
wavefunction. Also, what we obtained for $\beta^2$ values from 0.03 GeV$^2$ 
to 1 GeV$^2$ was that the proton spectrum results were fairly insensitive 
to the $\beta^2$ value.  
The resulting invariant spectrum of the recombined protons 
are shown by the coarsely-dotted line (typically for $\beta^2 =$ 0.495 GeV$^2$ but
the result doesn't change much for other $\beta^2$ values) in 
the bottom panel of Fig.~\ref{Fig:wf_check}. 
The recombined proton yield by the factorized wavefunction 
is significantly lower than the one by nonfactorizable wavefunction. 
This comparison indicates that it may be more significant in the baryon case
than in the meson case what form of the wavefunction is taken for the
prediction of the recombination process. Thus, the proper treatment of the wavefunction 
seems particularly important for the baryon production in the recombination process. 

We note that there are also other approaches for the recombination process: 
Refs. \cite{grec-1} and \cite{hwa-yang} considered 
soft-hard recombination with a covariant coalescence model 
and the fragmentation as a part of recombination, respectively.
In addition, a recent work by A. Majumder, E. Wang, and X.-N. Wang 
\cite{frag-re} discussed a derivation of the recombination model
from field theory description of jet fragmentation. 
They noted that an ad hoc formulation of the recombination model 
is only valid under some strict conditions on the hadron wavefunction.
Therefore, we note that the present development is not the first one 
to consider the dependence of hadron wavefunctions 
in the recombination process.

\subsection{Fragmentation and Jet Quenching}\label{frag}

Inclusive hadron production by fragmentation at large momentum transfer 
can be described well by perturbative quantum chromodynamics (pQCD). 
In the framework of pQCD, the invariant yield of hadron $h$ 
with momentum $P$ is given by \cite{fries}
\begin{equation}
E{{d^{3}N_{h}^{frag}}\over{dP^{3}}} = \sum_{a}\int_{0}^{1} {{dz}\over{z^{2}}}
D_{a \rightarrow h}(z) E_{a}{{d^{3} N_{a}^{pert}}\over{dp_{a}^{3}}},
\label{fragmentation}
\end{equation}
where the sum runs over all constituent quark species $a$ in $h$.
For the spectrum of parton $a$ with momentum $p_{a} = P/z$ at midrapidity, 
we use the parameterization by pQCD: 
\begin{equation}
E_{a}{{d^{3}N_{a}^{pert}}\over{dp_{a}^{3}}} 
= {{d^{2} N_{a}^{pert}}\over{2\pi p_{aT}dp_{aT}dy}}|_{y=0}
= {{K}\over{\pi}} {{C}\over{(1+p_{aT}/B)^{\kappa}}},
\label{pqcd}
\end{equation}
where the parameters $C$, $B$, and $\kappa$ are taken from 
a leading order pQCD calculations \cite{sriva-1}.
$K$ = 1.5 is taken in order to consider higher order corrections 
approximately \cite{fries}. 
Note that the number of partons in different collision centralities 
are obtained by scaling the number of 
binary nucleon-nucleon collisions ($N_{coll}$) 
or, equivalently, by the nuclear thickness function ($T_{AA}$).
The probability that parton $a$ fragments into hadron $h$ is 
taken into account by the fragmentation function 
\begin{equation}
D_{a \rightarrow h}(z) = N z^{\gamma} (1 - z)^{\delta},
\label{dah}
\end{equation}
where the numerical values of $N$, $\gamma$, and $\delta$ are taken from 
the parameterization by Kniehl, Kramer, and P\"{o}tter 
(Tabel 2 of Ref. \cite{kkp} and the website 
"\url{http://www.desy.de/~poetter/kkp.f}") 
for fragmentation of pions, 
kaons, protons and antiprotons. We call this parametrization as the KKP parametrization. 
$\Lambda$ fragmentation function is taken from the work
by de Florian, Stratmann and Vogelsang (Tabel 1 of Ref. \cite{flori-1}).

Finally, the energy loss of energetic partons 
(so called jet quenching), especially, 
in central collisions is considered 
by the following parameterization \cite{baier-1,mueller-1}
\begin{equation}
\Delta p_{T}(b, p_{T}) = \epsilon(b) \sqrt{p_{T}} 
{{\langle L \rangle(b)}\over{R_{A}}},
\label{quenching}
\end{equation}
where $R_{A}$ is the radius of nucleus $A$, 
$\langle L \rangle(b)$ is the geometrical factor of 
the overlap zone of two nuclei, and $\epsilon(b)$ is 
the energy loss parameter of the hot medium with impact parameter $b$. 
The detailed functional forms of $\langle L \rangle(b)$ and 
$\epsilon(b)$ are the same as Ref.~\cite{fries}: 
\begin{equation}
\epsilon(b) = \epsilon_{0}{{1-\exp[-(2R_{A}-b)/R_{A}]}
\over{1-\exp(-2)}}
\label{epsilon}
\end{equation}
and
\begin{equation}
\langle L \rangle(b)={{\sqrt{R_{A}^{2}-(b/2)^{2}}+(R_{A}-b/2)} 
\over {2}},
\label{loverb}
\end{equation}
but, practically speaking, it is reasonable to assume that 
$\langle L \rangle(b) \simeq R_{A}$ and 
$\epsilon(b) \simeq \epsilon_{0}$ = 0.82 GeV$^{1/2}$ 
for the most central collisions as $b \rightarrow 0$.

\section{results}\label{result}

\subsection{Invariant Spectra}\label{inva}

\begin{figure}[t!]
\begin{center}
\includegraphics[width=13cm]{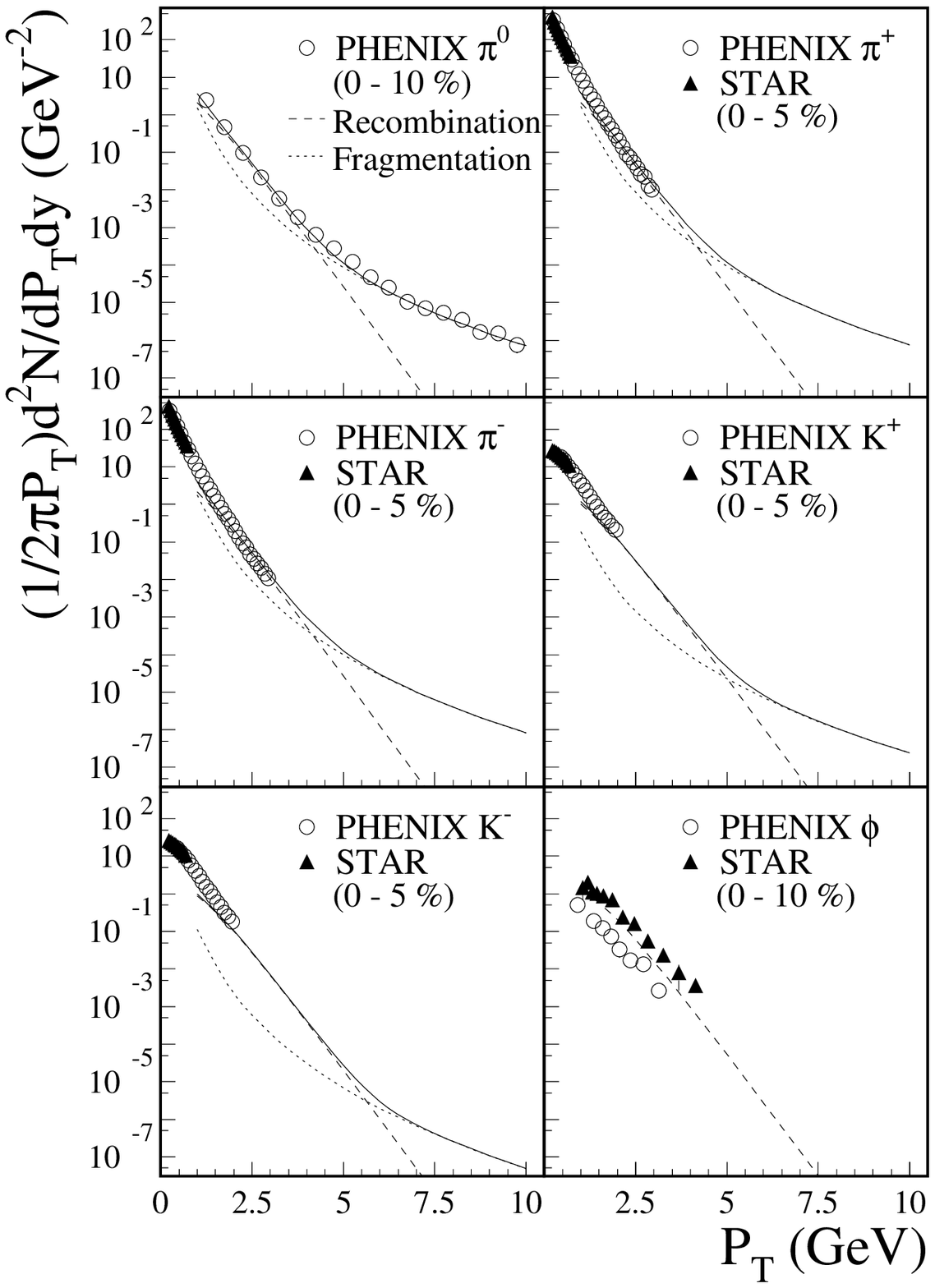}
\end{center}
\caption[]{Invariant spectra of mesons at midrapidity 
for central Au + Au collisions at $\sqrt{s_{NN}}$ =  200 GeV.
The dashed and dotted lines represent the model calculations 
from the recombination and the fragmentation, respectively.
The solid lines are the sum of two contributions.
The open circles are the published data by the PHENIX 
collaboration \cite{p1,p2,p3}, and the solid triangles 
are those by the STAR collaboration \cite{s1,s2}.}
\label{Fig:meson}
\end{figure}

Fig.~\ref{Fig:meson} shows the numerical results on 
the invariant spectra of various mesons at midrapidity 
for central Au + Au collisions at $\sqrt{s_{NN}}$ = 200 GeV. 
In Fig.~\ref{Fig:meson}, we compare our calculations 
for the meson spectra with the published PHENIX and STAR data 
\cite{p1,p2,p3,s1,s2} up to $P_{T}$ = 10 GeV in order to show 
the overall shapes, especially, the transition regions near 5 GeV. 
The neutral pion spectrum was measured by PHENIX up to 10 GeV in $P_{T}$,
but lacks data in a low $P_{T}$ region.
In contrary, the charged pions were measured only at low $P_{T}$ 
up to 3 GeV with high precision. 
However, the high $P_{T}$ spectra of charged pions are expected 
to be very similar to those of neutral pions. 
For charged kaons, PHENIX measured up to 2 GeV, 
and STAR measured up to about 0.7 GeV in $P_{T}$. 
In general, the data by PHENIX and STAR agree quite well 
in the overlapped phase space except $\phi$: 
the STAR data is about a factor of three larger than 
the PHENIX data.

In Fig.~\ref{Fig:meson}, the dashed and dotted lines 
represent the model calculations from the recombination 
and the fragmentation, respectively, and 
the solid lines are the sum of two contributions.
In the $\pi^{0}$ spectra, the two distinguished $P_{T}$ regions of 
hadron production are manifest.
Although the transient $P_{T}$ depends on the particle species,
the recombination process is dominant between $\sim$ 2 and 5 GeV,
and the fragmentation is dominant above 5 GeV.
Our calculation is not expected to reproduce the hadron spectra 
below about 2 GeV in $P_{T}$.
In such a very low $P_{T}$ region, the calculation underestimates 
the experimental data significantly, implying that 
other processes like the transverse flow,
the secondary decay of hadron resonances, and the binding energy effect 
become important. 
Our calculation reproduces the measured meson $P_{T}$ spectra 
larger than 2 GeV reasonable well, including the strange mesons.
Note that we do not plot the fragmentation contribution for $\phi$
due to the lack of the fragmentation function. 

\begin{figure}[t!]
\begin{center}
\includegraphics[width=16cm]{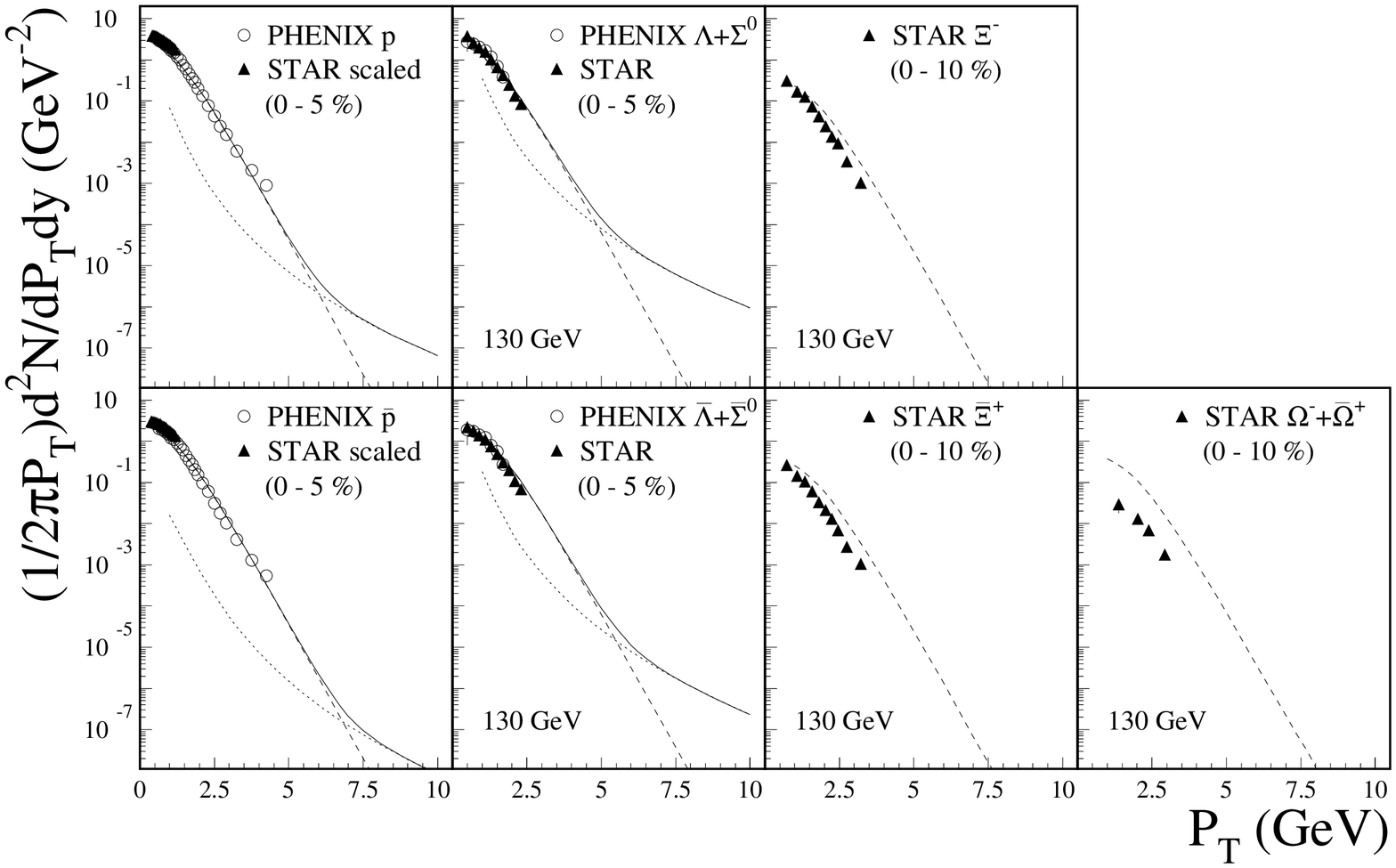}
\end{center}
\caption[]{Invariant spectra of baryons at midrapidity 
for central Au + Au collisions at $\sqrt{s_{NN}}$ =  200 GeV.
The dashed and dotted lines represent the model calculations 
from the recombination and the fragmentation, respectively.
The solid lines are the sum of two contributions.
The open circles are the published data by the PHENIX 
collaboration \cite{p2,p4} whereas the solid triangles 
are those by the STAR collaboration \cite{s1,s3,s4}. 
For a fair comparison between two sets of the data, the published 
$p$ and $\bar{p}$ spectra by the STAR collaboration \cite{s1}
are scaled by 0.6, which removes the contribution by 
the weak decays of $\Lambda$, $\Sigma^{0}$, and their antiparticles.
Note that the experimental invariant spectra of hyperons are 
for $\sqrt{s_{NN}}$ = 130 GeV, 
whereas all model calculations are for $\sqrt{s_{NN}}$ = 200 GeV.
}
\label{Fig:baryon}
\end{figure}

Fig.~\ref{Fig:baryon} shows the numerical results on 
the invariant spectra of baryons up to $P_T$ = 10 GeV at midrapidity 
for central Au + Au collisions at $\sqrt{s_{NN}}$ = 200 GeV. 
Fig.~\ref{Fig:baryon} also compares the calculations with 
the published experimental data with the open circles by PHENIX \cite{p2,p4} 
and the solid triangles by STAR \cite{s1,s3,s4}. 
The left most column of Fig.~\ref{Fig:baryon} is for protons and antiprotons; 
PHENIX and STAR measured up to about 5 and 1.2 GeV, respectively. 
Note that the published $P_{T}$ spectra for protons and antiprotons 
by STAR \cite{s1} are about 40 \% higher than those by PHENIX \cite{p2}. 
This difference comes from the fact that the contributions from 
the $\Lambda$ and $\Sigma^{0}$ decays are removed only for PHENIX. 
For a fair comparison between the two data sets, 
the $p$ and $\bar{p}$ spectra by the STAR collaboration 
are scaled by 0.6 in Fig.~\ref{Fig:baryon}. After scaling down, 
the STAR spectra agree quite well with the PHENIX spectra
in the overlapped phase space. The present model reproduces 
the measured proton and antiproton spectra reasonably well. 
The model also predicts that the transient $P_{T}$ for baryons 
from recombination to fragmentation is somewhat higher than that for mesons.

In Fig.~\ref{Fig:baryon}, the experimental invariant spectra 
of $\Lambda + \Sigma^{0}$, $\Xi^{-}$, $\Omega^{-}$, 
and their antiparticles are for $\sqrt{s_{NN}}$ = 130 GeV. 
But all model calculations are for $\sqrt{s_{NN}}$ = 200 GeV 
because all input parameters of the model calculations 
are available only for $\sqrt{s_{NN}}$ = 200 GeV.
Furthermore, due to the lack of the fragmentation functions,
we do not plot the fragmentation contribution for $\Xi$ and $\Omega$.
Because of the difference in beam energy, the model 
overestimates the yields of $\Lambda + \Sigma^{0}$, 
$\Xi^{-}$, $\Omega^{-}$, and their antiparticles, and the discrepancy 
is larger for a larger number of strange quarks in a given baryon. 

\subsection{Yield Ratios}\label{ratio}

\begin{figure}[t!]
\begin{center}
\includegraphics[width=13cm]{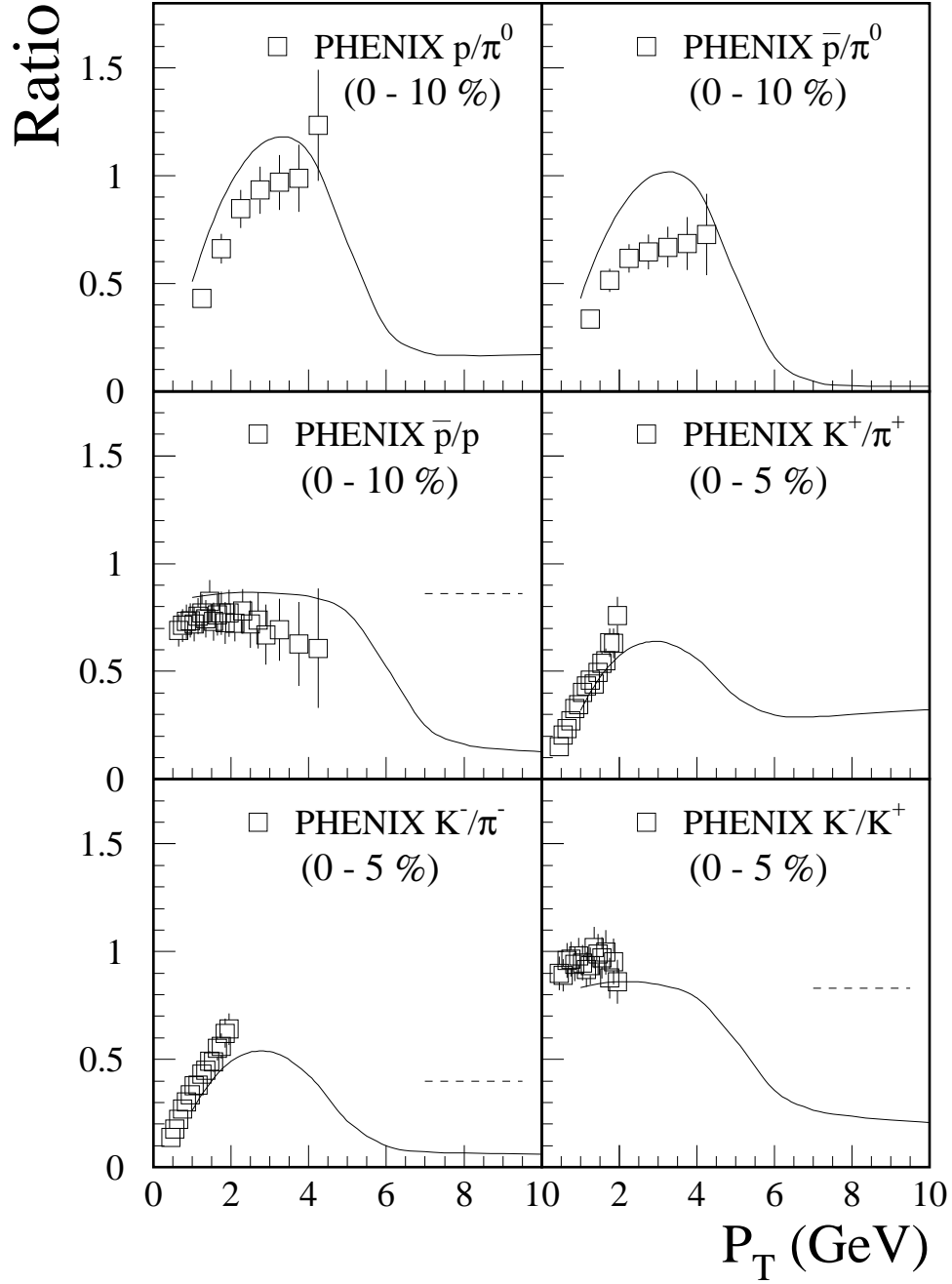}
\end{center}
\caption[]{Calculated hadron yield ratios (solid lines) 
as a function of $P_{T}$ in comparison with the PHENIX data \cite{p2}. 
For the comparison, we also show the model calculations 
by Fries {\it et al.}, in $\bar{p}/p$, $K^{-}/\pi^{-}$, 
and $K^{-}/K^{+}$ by dashed lines \cite{fries}.}
\label{Fig:ratio}
\end{figure}

One of the most interesting data from RHIC is the yield ratio of 
protons(or antiprotons) to pions at the intermediate transverse momentum 
region (2 $< P_{T} <$ 5 GeV) in central heavy-ion collisions. 
The $p/\pi$ and $\bar{p}/\pi$ ratios rise steeply with $P_{T}$ 
up to about 2.5 GeV, but levels off at about 1 and 0.7, respectively,
in 2.5 $< P_{T} <$ 5 GeV for the most central 
10 \% Au + Au collisions \cite{p2,p5}. 
At $P_{T} >$ 2 GeV, $p/\pi$ and $\bar{p}/\pi$ for peripheral collisions 
are similar to those for elementary $p + p$ and $e^{+} e^{-}$ collisions,
and the ratios increase from peripheral to central collisions. 
Since the hydrodynamic model, which had been rather successful 
in describing the low $P_{T}$ hadron spectra, 
could not explain the centrality dependence of limiting values, 
a recombination mechanism of hadronization at intermediate $P_{T}$ 
was proposed as a possible resolution \cite{fries, grec-1, hwa-yang, vg, hy}. 
The recombination process naturally explains that 
the $p/\pi$ and $\bar{p}/\pi$ ratios level off in 2 $< P_{T} <$ 5 GeV, 
and fall sharply near $P_{T} \simeq$ 5 GeV where 
the fragmentation takes over the recombination. 
Similar trends can also be found in the present calculation. 
The top row of Fig.~\ref{Fig:ratio} shows the results from our calculations  
for the $p/\pi^{0}$, $\bar{p}/\pi^{0}$, and $\bar{p}/p$ ratios 
in comparison with the published PHENIX data \cite{p2}. 
As $P_{T}$ increases, the $p/\pi^{0}$ and $\bar{p}/\pi^{0}$ ratios 
rises, reach the maximum values around 3 GeV, 
decrease sharply, and, finally, become constant 
at about 0.1 for $P_{T} >$ 6 GeV. 
In addition, the $\bar{p}/p$ ratio is almost constant at about 0.9 
for $P_{T} <$ 5 GeV. However, it also decreases with $P_{T}$, 
and become almost constant at about 0.1 for $P_{T} >$ 7 GeV,
which is very different from the previous calculation by 
Fries {\it et al.} (the dashed line in Fig.~\ref{Fig:ratio}) \cite{fries}. 
Although there is no dispute on KKP parametrizations~\cite{kkp} for the gluon
fragmentation, the quark fragmentation is a problem because the KKP
fragmentation functions are not fully flavor seperated and one has to
make additional assumptions to seperate contributions from different
flavors. Incidentally, the recent STAR data on identified hadrons~\cite{star2006}
reveal a poor job of the KKP fragmentation functions for $p$ and ${\bar p}$ yields,
possibly due to the lack of flavor seperation.
It seems that the sea quark contributions in $K^\pm$ and 
${\bar p}/p$ are particularly problematic. For instance, as stated in Ref.~\cite{kkp},
the $d$ quark in $K^\pm$ does not behave sea-like contrary to expectations.
Thus, even a slight difference in handling sea quark contributions could
make a large difference in the predictions of ratios for $K^-/K^+$
and ${\bar p}/p$. Our results are based on maintaining expected smallness
of sea quark contributions consistently not only in the pion case but also
in other hadron cases. Our low ratio for $P_{T} >$ 7 GeV in ${\bar p}/p$ of
Fig.~\ref{Fig:ratio} is consequently due to the dominance of valence contributions. 
It is a fact that incident heavy ions possess valence quarks, but not antiquarks.
In other words, the charge conjugation symmetry is 
already broken in RHIC environment due to the initial nuclei 
carry only nucleons (not antinucleons). 
Since the baryon number (or, equivalently, the quark number) 
must be conserved throughout the reactions, more protons than 
antiprotons are expected in the fragmentation region. 

For more comparisons on the hadron yield ratios, 
the bottom row of Fig.~\ref{Fig:ratio} shows $K^{+}/\pi^{+}$, 
$K^{-}/\pi^{-}$, and $K^{-}/K^{+}$. 
Although the measured $P_{T}$ range of $K^{\pm}$ is limited, 
the present estimates are in reasonable agreements with the data. 
The $K^{+}/\pi^{+}$ and $K^{-}/\pi^{-}$ ratios increase with $P_{T}$, 
and reach maximum around $P_{T}$ =  3 GeV. 
If $P_{T}$ further increases, $K^{+}/\pi^{+}$ and $K^{-}/\pi^{-}$ 
decrease, and level off at some constants. 
The $P_{T}$ dependence of the $K^{-}/K^{+}$ ratio is very similar 
to that of  $\bar{p}/p$. Especially, we note that 
the present results on $K^{-}/\pi^{-}$ and $K^{-}/K^{+}$ 
at high $P_{T}$ region, where the fragmentation is dominant, 
are quite different from the previous model calculations 
by Fries {\it et al.} \cite{fries}.
As discussed above, even a slight difference in handling sea quark 
contributions could make a large difference in the predictions of ratios 
for $K^-/K^+$ as well as $K^-/\pi^-$.
The forthcoming RHIC data at high $P_{T}$, e.g., 
the PHENIX data with newly installed aerogel detector, 
may help to further clarify the flavor seperation issue in the KKP 
fragmentation functions~\cite{kkp}.

\subsection{Nuclear Modification Factors}\label{rcp}

Another important feature of the RHIC data can be identified in 
the nuclear modification factor $R_{CP}$, which is defined by 
the $N_{coll}$ scaled central to peripheral yield ratios: 
\begin{equation}
R_{CP} = {{Yield^{central}/<N_{coll}^{central}>}\over
{Yield^{peripheral}/<N_{coll}^{peripheral}>}}.
\label{Eq:rcp}
\end{equation}
The RHIC experiments observed that the $R_{CP}$ parameters of 
various mesons in $P_{T} >$ 2 GeV in central collisions were suppressed 
with respect to the $N_{coll}$ scaled $p + p$ and peripheral collision data.
Moreover, the suppression in the intermediate transverse momentum region 
(2 $< P_{T} <$ 4 GeV) was only for mesons, but not for baryons. 
The experimental $R_{CP}$ parameter of protons in intermediate 
transverse momentum region is unity, 
which is completely consistent with $N_{coll}$ scaling. 
The $R_{CP}$ of $\Lambda$ and $\bar{\Lambda}$ are also close to unity
in an intermediate $P_{T}$ region, but somewhat smaller than protons.

\begin{figure}[t!]
\begin{center}
\includegraphics[width=8.5cm]{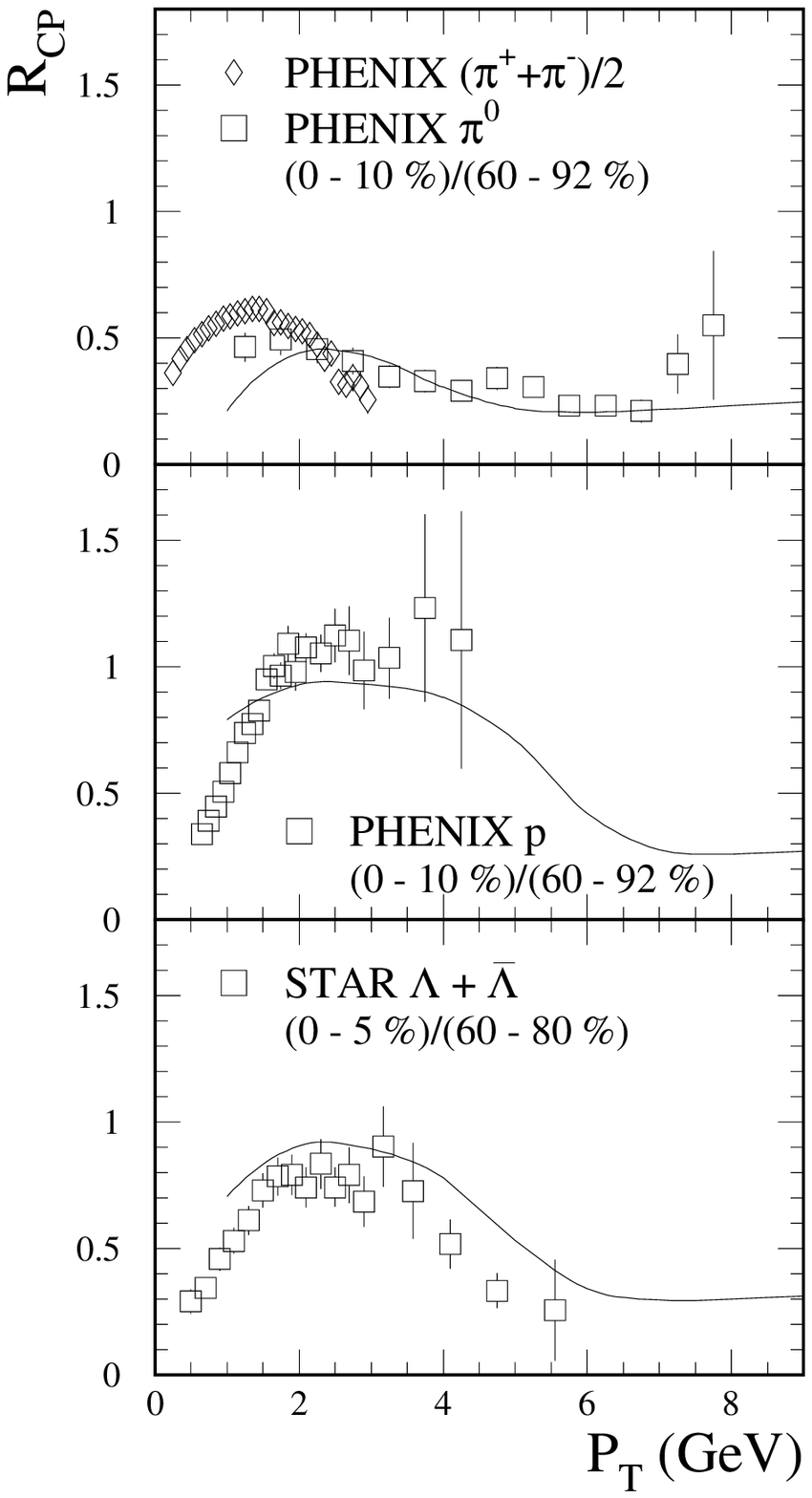}
\end{center}
\caption[]{Nuclear modification factors $R_{CP}$ as a function of $P_{T}$ 
for $\pi$, $p$, and $\Lambda + \bar{\Lambda}$.
The solid lines are from the present calculation. 
}
\label{Fig:rcp}
\end{figure}

One of possible explanations for the suppression of 
hadron yields at high $P_{T}$ and a distinguished behavior 
of mesons and baryons at the intermediate $P_{T}$ region is 
the combined effect of recombination and fragmentation. 
Fig.~\ref{Fig:rcp} shows the estimated $R_{CP}$ parameters 
of $\pi$, $p$, and $\Lambda + \bar{\Lambda}$ as a function of $P_{T}$. 
For pions, we plot charged and neutral pions together, 
as almost no difference is expected from the present model. 
In the present model calculation, we scaled the hadron yields 
due to the fragmentation by the number of binary collisions. 
For the recombination part, 
the ratio of $V \displaystyle{\prod_{a}}\gamma_{a}$ 
in Eq.(\ref{extend-spectrum}) for peripheral collisions 
to that in central collisions was assumed 
as 40 \% of the number of participant ($N_{part}$) ratio:
\begin{equation}
{{(V\displaystyle{\prod_{a}}\gamma_{a})^{peripheral}}\over
{(V\displaystyle{\prod_{a}\gamma_{a}})^{central}}} = c_{1}~~
{{N_{part}^{peripheral}}\over{N_{part}^{central}}},
\label{volume}
\end{equation}
where $c_1$ = 0.4 fitted by the $R_{CP}$ parameters of 
$\pi$, $p$, and $\Lambda + \bar{\Lambda}$, simultaneously, 
by fixing the temperature at 175 MeV as it is almost 
independent of the collision centrality \cite{fwang}. 
Since the quark fugacities are also almost constant 
except for the very peripheral collisions \cite{fwang}, 
the factor $c_1$ mostly reflects the effect of volume. 
As a result, the fact that $c <$ 1 is understandable as the 
flow velocity is larger for more central collisions. 
The agreement between the present calculations and the experimental data 
are reasonable for all considered hadron species.

\section{conclusions}\label{conc}

We have presented an extended formalism of the recombination model 
to analyze the effects from the variation of 
the hadron's light-front wavefunctions. 
Two different functional forms of the light-front wavefunction, 
which are the Gaussinan form and the power-law form, are tested in detail. 
The hadron spectra are indeed sensitive to the shape of the wavefunctions.
However, when we fit the wavefunction parameters to the physical 
observables, such as the average charge radius, 
the final spectra are very similar each other. 
We discuss our numerical results in comparison with the published RHIC data, 
especially, from the PHENIX and STAR collaborations. 
In the hadron spectra, the recombination of thermal partons dominates 
at the intermediate transverse momentum region between 2 and 5 GeV, 
and the fragmentation dominates at high $P_{T}$ larger than 5 GeV. 
The yield ratios and the nuclear modification factors for various hadron 
species are also estimated. 
In general, the present model, which combines the recombination and 
fragmentation processes, are quite consistent with the experimental data. 
We have also discussed new predictions on $\bar{p}/p$ and $K^{-}/K^{+}$ ratios, 
including the jet quenching effects to the fragmentation mechanism.

\begin{acknowledgments}
This work is supported in part by a grant from the U.S. Department of 
Energy(DE-FG02-96ER40947) and a brain pool program from the KOFST.
CRJ thanks to the faculty and staff at the School of Physics at Seoul 
National University for the hospitality during the Sabbatical visit
while this work was made. We would like to thank B. Mueller, R. Fries, S. Bass and
X.-N. Wang for useful discussion and information. The National Energy
Research Scientific Computer Center is also acknowledged for the grant
of supercomputing time.
\end{acknowledgments}


\begin{table}[t!]
\caption{The deduced $\beta^2$ for the Gaussian LF wavefunctions 
used in this paper and the corresponding average 
charge radius square for various hadrons.}
\begin{center}
\begin{tabular}{ccc} \\ \hline
Particle  & $\beta^2$ (GeV$^2$) & $<r^{2}>$ (fm$^2$) \\ \hline
$\pi$ & 0.825 & 0.44 \\
$K$ & 1.06 & 0.34 \\
$\phi$ & 1.02 & 0.34 \\
$p$ & 0.495 & 0.76 \\
$\Lambda$ & 0.45 & 0.76 \\
$\Xi$ & 0.47 & 0.76 \\
$\Omega$ & 0.48 & 0.76 \\ \hline
\end{tabular}
\end{center}
\label{Tab:beta2}
\end{table}

\newpage

\begin{thebibliography}{99}
\bibitem{lee1} T. D. Lee, Nucl. Phys. {\bf A750}, 1 (2005).
\bibitem{b-wp} I. Arsene {\it et al.}, BRAHMS Collaboration, 
 Nucl. Phys. {\bf A757}, 1 (2005). 
\bibitem{p-wp} K. Adcox {\it et al.}, PHENIX Collaboration, 
 Nucl. Phys. {\bf A757}, 184 (2005). 
\bibitem{ph-wp} B. B. Back {\it et al.}, PHOBOS Collaboration, 
 Nucl. Phys. {\bf A757}, 28 (2005).
\bibitem{s-wp} J. Adams {\it et al.}, STAR Collaboration, 
 Nucl. Phys. {\bf A757}, 102 (2005). 
\bibitem{wang-1} X. N. Wang,
 Phys. Rev. C {\bf 63}, 054902 (2001).
\bibitem{wang-2} M. Gyulassy, I. Vitev, and X. N. Wang, 
 Phys. Rev. Lett. {\bf 86}, 2537 (2001).
\bibitem{fries} R. J. Fries, B. M\"{u}ller, C. Nonaka, and S. A. Bass,
 Phys. Rev. C {\bf 68}, 044902 (2003); 
 Phys. Rev. Lett. {\bf 90}, 202303 (2003).
\bibitem{mola-1} D. Moln\'{a}r and S. Voloshin,
 Phys. Rev. Lett. {\bf 91}, 092301 (2003).
\bibitem{grec-1} V. Greco, C. M. Ko, and P. Levai,
 Phys. Rev. C {\bf 68}, 034904 (2003).
\bibitem{qc-1} T. S. Biro, P. Levai, and J. Zimanyi,
 Phys. Lett. B {\bf 347}, 6 (1995).
\bibitem{qc-2} P. Csizmadia and P. Levai,
 J. Phys. G {\bf 28}, 1997 (2002).
\bibitem{hong-1} B. Hong,
 J. Korean Phys. Soc. {\bf 45}, L795 (2004).
\bibitem{pbm-1} A. Andronic, P. Braun-Munzinger, K. Redlich, and J. Stachel,
 Phys. Lett. B {\bf 571}, 36 (2003). 
\bibitem{torri} G. Torrieri and J. Rafelski,
 J. Phys. Conf. Ser. {\bf 5}, 246 (2005); arXive:hep-ph/0409160.
\bibitem{lqcd} F. Karsch, Nucl. Phys. {\bf A698}, 199 (2002).
\bibitem{ji-1} H.-M. Choi and C.-R. Ji,
 Phys. Rev. D {\bf 56}, 6010 (1997).
\bibitem{mass-m} S. Godfrey and N. Isgur, 
 Phys. Rev. D {\bf 32}, 189 (1985). 
\bibitem{mass-b} S. Capstick and N. Isgur, 
 Phys. Rev. D {\bf 34}, 2809 (1986).
\bibitem{mass-3} H.-M. Choi and C.-R.Ji,
 Phys. Rev. D {\bf 59}, 074015 (1999). 
\bibitem{ji-2} C.-R. Ji and S. R. Cotanch,
 Phys. Rev. D {\bf 41}, 2319 (1990).
\bibitem{schl-1} F. Schlumpf, Phys. Rev. D {\bf 50}, 6895 (1994).
\bibitem{hwa-yang} R. C. Hwa and C. B. Yang, 
 Phys. Rev. C {\bf 70}, 024905 (2004). 
\bibitem{frag-re} A. Majumder, E. Wang, and X.-N. Wang, 
 arXiv:nucl-th/0506040.
\bibitem{sriva-1} D. K. Srivastava, C. Gale, and R. J. Fries,
 Phys. Rev. C {\bf 67}, 034903 (2003).
\bibitem{kkp} B. A. Kniehl, G. Kramer, and B. P\"{o}tter, 
 Nucl. Phys. {\bf B582}, 514 (2000).
\bibitem{flori-1} D. de Florian, M. Stratmann, and W. Vogelsang, 
 Phys. Rev. D. {\bf 57}, 5811 (1998).
\bibitem{baier-1} R. Baier, Y. L. Dokshitzer, A. H. M\"{u}ller, and D. Schiff,
 J. High Energy Phys. {\bf 0109}, 033 (2001).
\bibitem{mueller-1} B. M\"{u}ller, 
 Phys. Rev. C {\bf 67}, 061901 (2003).
\bibitem{p1} S. S. Adler {\it et al.}, PHENIX Collaboration, 
 Phys. Rev. Lett. {\bf 91}, 072301 (2003). 
\bibitem{p2} S. S. Adler {\it et al.}, PHENIX Collaboration, 
 Phys. Rev. C {\bf 69}, 034909 (2004). 
\bibitem{p3} S. S. Adler {\it et al.}, PHENIX Collaboration, 
 Phys. Rev. C {\bf 72}, 014903 (2005). 
\bibitem{s1} J. Adams {\it et al.}, STAR Collaboration, 
 Phys. Rev. Lett. {\bf 92}, 112301 (2004). 
\bibitem{s2} J. Adams {\it et al.}, STAR Collaboration, 
 Phys. Lett. B {\bf 612}, 181 (2005). 
\bibitem{p4} K. Adcox {\it et al.}, PHENIX Collaboration, 
 Phys. Rev. Lett. {\bf 89}, 092302 (2002). 
\bibitem{s3} C. Adler {\it et al.}, STAR Collaboration, 
 Phys. Rev. Lett. {\bf 89}, 092301 (2002). 
\bibitem{s4} J. Adams {\it et al.}, STAR Collaboration, 
 Phys. Rev. Lett. {\bf 92}, 182301 (2004). 
\bibitem{p5} S. S. Adler {\it et al.}, PHENIX Collaboration, 
 Phys. Rev. Lett. {\bf 91}, 172301 (2003). 
\bibitem{vg} I. Vitev and M. Gyulassy, Phys. Rev. C {\bf 65}, 041902 (2002).
\bibitem{hy} R. C. Hwa and C. B. Yang, Phys. Rev. C {\bf 67}, 034902 (2003).
\bibitem{star2006} J. Adams {\it et al.}, STAR Collaboration, arXiv:nucl-ex/0601033.
\bibitem{fwang} F. Wang, Overview talk for the STAR Collaboration 
 given in Quark Matter 2005 Conference, Budapest, Hungary, August 4-9, 2005;
 Nucl. Phys. {\bf A}, (2006) in print.
\end{thebibliography}
\end{document}